\documentclass[multphys,vecphys]{svmult}
\usepackage{makeidx}  
\usepackage{graphicx} 
\usepackage{multicol}        
\usepackage[bottom]{footmisc}
\makeindex
\begin{document}

\title*{Multi-colour Imaging of Ultra-compact Objects in the Fornax Cluster}
\author{A. M. Karick\inst{1}\and M. D. Gregg\inst{1}\and
M. J. Drinkwater\inst{2}\and M. Hilker\inst{3}\and P. Firth\inst{2}}
\institute{University of California, Davis, USA,
\texttt{akarick@igpp.ucllnl.org} \and University of Queensland,
Australia \and Sternwarte der Universitat Bonn, Germany } \maketitle
Ultra-compact dwarf galaxies (UCDs) are a new type of galaxy we have
discovered in the central region of the Fornax and Virgo
clusters. Unresolved in ground-based imaging, UCDs have spectra
typical of old stellar systems. Ninety-two have been found in Fornax
\cite{Drink2004}\cite{Mieske2004}, making them {\it the most numerous
galaxy type in the cluster}.  Although they form a cluster wide
population, an over-density surrounds the central cluster galaxy,
NGC1399, fueling controversy over their nature and origin. Several
formation scenarios have been proposed. UCDs may be the remnant nuclei
of tidally stripped dE,N galaxies \cite{Bekki2001} or they may be the
bright tail of the globular cluster (GC) population associated with
NGC1399 [3,{\small ``UCOs''}]. Alternatively they may be the first
spectroscopically confirmed intracluster globular clusters (IGCs) in
Fornax, resulting from hierarchical star cluster formation
\cite{Fellhauer2002} and merging in intracluster space. The 5
brightest Fornax UCDs have M/L ratios indicating at lest some dark
matter, unlike typical GCs.

\subsubsection{Multicolour Imaging of the Fornax Cluster }
\label{sec:col}
We obtained deep multicolour
(u{\footnotesize$^\prime$}g{\footnotesize$^\prime$}r{\footnotesize$^\prime$}i{\footnotesize$^\prime$}z{\footnotesize$^\prime$})
imaging of the central region of the Fornax Cluster using the CTIO 4m
Mosaic Telescope. Fig. 1 ({\sc left}) shows the radial distribution of
the present population of UCDs, GCs and dE,Ns. Sixty-two UCDs were
discovered in our 2dF spectroscopic surveys of Fornax. An additional
30 were discovered in our recent VLT survey, from which candidates
were pre-selected using this multicolour data
\cite{Karick2005}. Fig. 1 ({\sc right}) shows the
g{\footnotesize$^\prime$}$-$i{\footnotesize$^\prime$} colour-magnitude
relation (CMR) for the UCDs, UCOs \cite{Mieske2004}, NGC1399 GCs
\cite{Dirsch2004}\cite{Kissler1998}\cite{Minniti1998}, and cluster
dwarf galaxies \cite{Ferguson1989}. All objects are spectroscopically
confirmed cluster members.

Bright UCDs and dEs follow similar CMRs: both populations become
redder with increasing luminosity. The slope for the CMR for bright
UCDs and fainter UCD candidates is $-$0.05~mag, similar to the slope
of the CMR of candidate UCDs ($-$0.07~mag) in Abell 1689
\cite{Mieske2004b}. This is qualitatively consistent with UCDs being
the stripped nuclei of dE,Ns. UCDs have similar colours to NGC1399
GCs, however they exhibit a larger spread in colour at 18$<$
i{\footnotesize$^\prime$}$<$20, possibly reflecting a mixed
metallicity or age spread. These results will be discussed in more
detail in a forthcoming paper.

At faint magnitudes, GC and UCDs cannot be distinguished by colour
alone. High resolution spectroscopy to measure their internal velocity
dispersions and metallicities, is needed to distinguish between GCs
and UCDs. If dark matter is required to explain their dynamics then
these faint compact objects are real galaxies or UCDs.

\vspace{-5mm}
\begin{figure}
\includegraphics[scale=0.29]{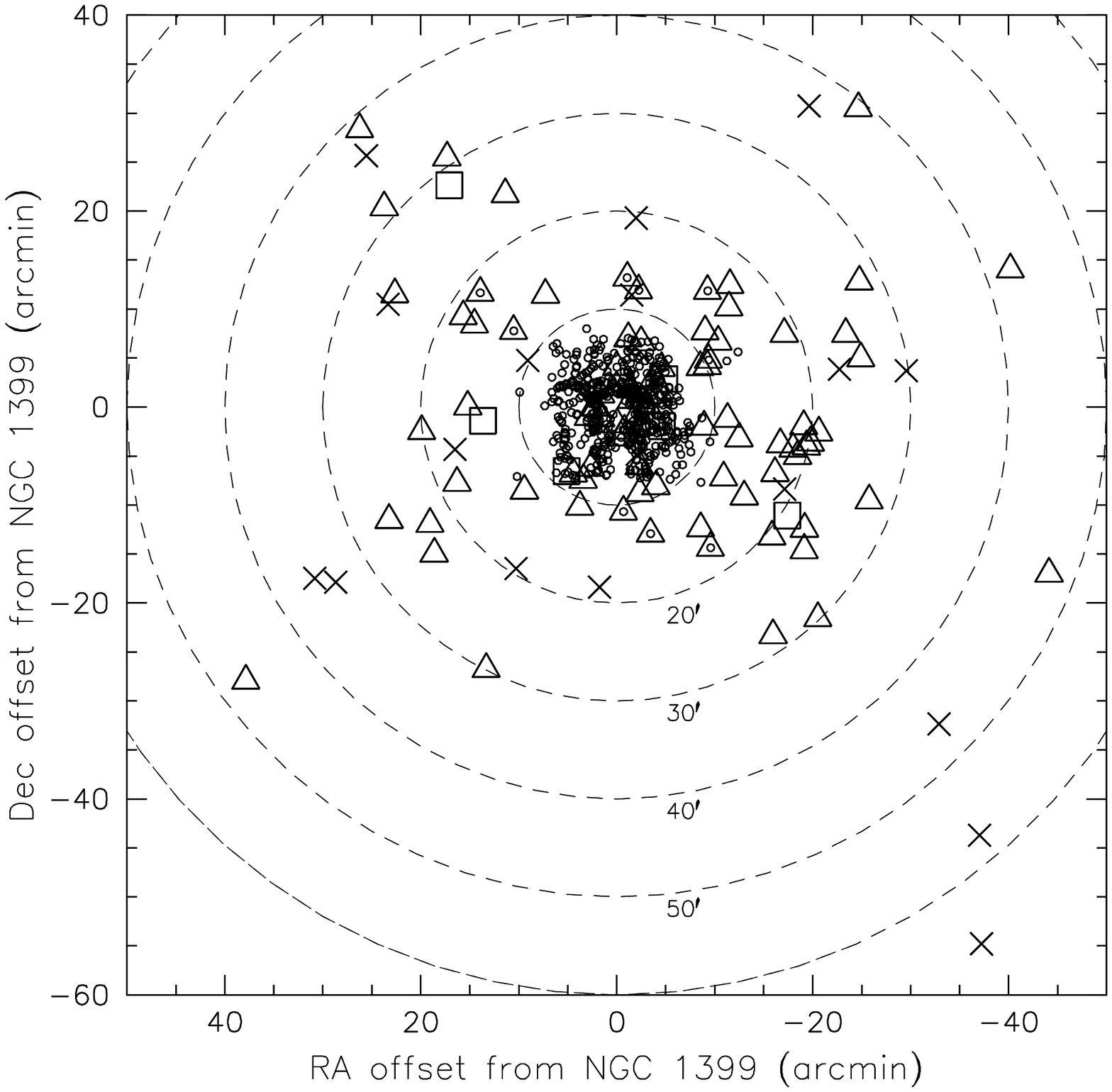}
\includegraphics[scale=0.29]{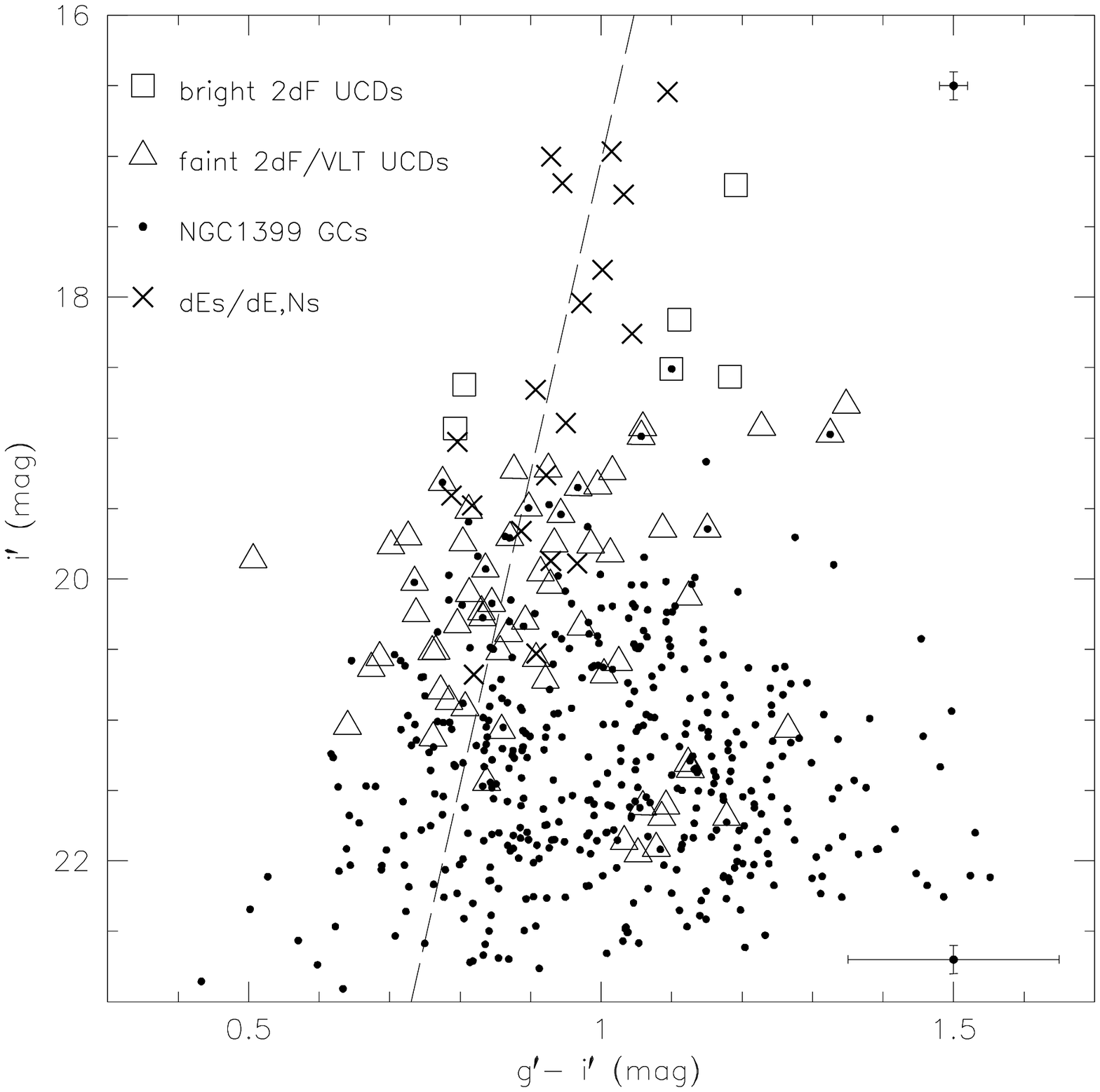}
\vspace{-5mm}
\caption{{\sc left:} Distribution of compact objects surrounding
NGC1399: bright 2dF UCDs (boxes), fainter 2dF/VLT UCDs (triangles),
dE,Ns (crosses), NGC1399 GCs and UCOs (small circles).  The NGC1399
GCs extend out to 10{\footnotesize$^\prime$} (survey limited) and
overlap a small fraction (7\%) of the innermost UCDs. A significant
number of the 2dF/VLT UCDs (25\%) overlap the UCOs which extend out to
20{\footnotesize$^\prime$}. {\sc right:}
g{\footnotesize$^\prime$}$-$i{\footnotesize$^\prime$} CMR for compact
objects in Fornax: 5 brightest 2dF UCDs (boxes), fainter 2dF/VLT UCDs
(triangles), cluster dEs and dE,Ns (crosses) and NGC 1399 globular
clusters (points). The dashed line shows the fit to the dwarf galaxy
CMR. Typical error bars for i{\footnotesize$^\prime$}$\sim$16.5 mag
and i{\footnotesize$^\prime$}$\sim$22.5 mag objects are also shown. }
\end{figure}

\vspace{-3mm} This project was supported by grants from the ARC and
ANSTO Access to Large Research Facilities scheme. This material is
based on part upon work supported by NSF grant no. 0407445 and carried
out at IGPP under the auspices of the US DOE by LLNL under contract
no. W-7405-Eng-48.

\printindex
\end{document}